\documentclass[12pt]{article}

\newcommand{\D}{D} 
\newcommand{\hD}{h} 
\newcommand{\K}{K}
\newcommand{\KK}{\hat K}
\newcommand{\hKG}{h}

\newcommand{\mbf}[1]{{\boldsymbol {#1} }}

\newcommand{\cU}{{\cal U}}
\newcommand{\cF}{{\cal F}}
\newcommand{\cH}{{\cal H}}
\newcommand{\cA}{{\cal A}}
\newcommand{\cN}{{\cal N}}
\newcommand{\cE}{{\cal E}}
\newcommand{\cM}{{\cal M}}

\newcommand{\vx}{{\bf x}}
\newcommand{\vy}{{\bf y}}
\newcommand{\vp}{{\bf p}}
\newcommand{\vA}{{\bf A}}
\newcommand{\valpha}{{\mbf \alpha}}

\newcommand{\ii}{{\rm i}}
\newcommand{\dd}{{\rm d}}

\newcommand{\R}{{\mathbb R}}
\newcommand{\C}{{\mathbb C}}
\newcommand{\Z}{{\mathbb Z}}

\newcommand{\Ref}[1]{(\ref{#1})}

\newcommand{\eq}{\begin{equation}}
\newcommand{\eqend}{\end{equation}}
\newcommand{\eqa}{\begin{eqnarray}}
\newcommand{\nonueqa}{\begin{eqnarray*}}
\newcommand{\eqaend}{\end{eqnarray}}
\newcommand{\nonueqaend}{\end{eqnarray*}}

\newcommand{\bma}[1]{\begin{array}{#1}}
\newcommand{\ema}{\end{array}}
\newcommand{\bc}{\begin{center}}
\newcommand{\ec}{\end{center}}

\usepackage{graphicx}
\usepackage{theorem,amssymb,amsbsy,latexsym}
\begin{document}
\pagestyle{myheadings}
%\markright{Interactions with external fields}
\markright{Bosons and fermions in external fields}

\parindent 0mm
\parskip 6pt

%\title{Interactions with external fields}
\title{Bosons and fermions in external fields}
\author{Edwin Langmann\\Mathematical Physics\\Physics
KTH\\AlbaNova\\SE-106 91 Stockholm\\Sweden\\E-mail:
langmann@theophys.kth.se}

\date{\today}

%Abstract: Contribution to the Encyclopedia of Mathematical Physics
%(Elsevier, 2006): a brief and (hopefully) pedagogical introduction to
%quantum field theory models describing particles in external fields is
%presented. Following the instructions, the only references are a few
%recommendations for further reading in the final section.

\maketitle

\section{Introduction}
In this article we discuss quantum theories which describe systems of
non-distinguish\-able particles interacting with external fields. Such
models are of interest also in the non-relativistic case (in quantum
statistical mechanics, nuclear physics, etc.), but the relativistic
case has additional, interesting complications: in the latter case
they are genuine quantum field theories, i.e.\ quantum theories with
an infinite number of degrees of freedom, with non-trivial features
like divergences and anomalies. Since interparticle interactions are
ignored, such models can be regarded as a first approximation to more
complicated theories, and they can be studied by mathematically
precise methods.

Models of relativistic particles in external electromagnetic fields
have received considerable attention in the physics literature, and
interesting phenomena like the Klein paradox or particle-antiparticle
pair creation in over-critical fields have been studied; see
\cite{RFK} for an extensive review.  We will not discuss these physics
questions but only describe some proto-type examples and a general
Hamiltonian framework which has been used in mathematically precise
work on such models. The general framework for this latter work is the
mathematical theory of Hilbert space operators (see e.g.\ \cite{RS}),
but in our discussion we try to avoid presupposing knowledge of that
theory.  As shortly mentioned in the end, this work has had close
relations to various topics of recent interest in mathematical
physics, including anomalies, infinite dimensional geometry and group
theory, conformal field theory, and noncommutative geometry.

We restrict our discussion to spin-$0$ bosons and spin-$\frac12$
fermions, and we will not discuss models of particles in external
gravitational fields but only refer the interested reader to
\cite{DeWitt}.  We also only mention in passing that external field
problems have been also studied using functional integral approaches,
and mathematically precise work on this can be found in the extensive
literature on determinants of differential operators.

\section{Examples} Consider the {\em Schr\"odinger equation}
describing a non-relativistic particle of mass $m$ and charge $e$ moving in
three dimensional space and interacting with an external vector- and scalar
potential $\vA$ and $\phi$,
\eq \ii\partial_t\psi = H\psi,\quad H = \frac1{2m}(-\ii \nabla +
e\vA)^2 -e\phi \label{Sch} \eqend
(we set $\hbar=c=1$, $\partial_t=\partial/\partial t$, and $\psi$,
$\phi$ and $\vA$ can depend on the space and time variables $\vx\in
\R^3$ and $t\in\R$).  This is a standard quantum mechanical model,
with $\psi$ the one-particle wave function allowing for the usual
probabilistic interpretation. One interesting generalization to the
relativistic regime is the {\em Klein-Gordon equation}
\eq \left[\left(\ii {\partial_t} +e\phi\right)^2 - (
  -\ii\nabla + e\vA)^2 - m^2 \right]\psi  = 0 \label{KG} 
\eqend
with a $\C$-valued function $\psi$. There is another important relativistic
generalization, the {\em Dirac equation}
\eq \left[ \left(\ii{\partial_t}+ e\phi\right) - (-\ii \nabla +
e\vA)\cdot \valpha + m  \beta \right]\psi=0 \label{D} \eqend
with $\valpha=(\alpha_1,\alpha_2,\alpha_3)$ and $\beta$ hermitian $4\times
4$ matrices satisfying the relations
\eq \alpha_i\alpha_j+ \alpha_j\alpha_i =\delta_{ij} , \quad \alpha_i\beta =
-\beta\alpha_i,\quad \beta^2 = 1  \eqend
and a $\C^4$-valued function $\psi$ (we write $1$ also for the
identity). These two relativistic equations differ by the
transformation properties of $\psi$ under Lorentz transformations: in
\Ref{KG} it transforms like a scalar and thus describes spin-$0$
particles, and it transforms like a spinor describing spin-$\frac12$
particles in \Ref{D}. While these equations are natural relativistic
generalizations of the Schr\"odinger equation, they no longer allow to
consistently interpret $\psi$ as one-particle wave functions. The
physical reason is that, in a relativistic theory, high energy
processes can create particle-antiparticle pairs, and this makes the
restriction to a fixed particle number inconsistent.  This problem can
be remedied by constructing a many-body model allowing for an
arbitrary number of particles and anti-particles. The requirement that
this many-body model should have a groundstate is an important
ingredient in this construction.

It is obviously of interest to formulate and study many-body models of
non-distinguishable already in the non-relativistic case.  An
important empirical fact is that such particles come in two kinds,
{\em bosons} and {\em fermions}, distinguished by their exchange
statistics (we ignore the interesting possibility of exotic
statistics). For example, the fermion many-particle version of
\Ref{Sch} for suitable $\phi$ and $\vA$ is a useful model for
electrons in a metal.  An elegant method to go from the one- to the
many-particle description is the formalism of {\em second
quantization}: one promotes $\psi$ to a quantum field operator with
certain (anti-) commutator relations, and this is a convenient way to
construct the appropriate many-particle Hilbert space, Hamiltonian,
etc. In the non-relativistic case, this formalism can be regarded as
an elegant reformulation of a pedestrian construction of a many-body
quantum mechanical model, which is useful since it provides convenient
computational tools. However, this formalism naturally generalizes to
the relativistic case where the one-particle model no longer has an
acceptable physical interpretation, and one finds that one can
nevertheless can give a consistent physical interpretation to \Ref{KG}
and \Ref{D} provided that $\psi$ are interpreted as quantum field
operators describing bosons and fermions, respectively. This
particular exchange statistics of the relativistic particles is a
special case of the {\em spin-statistics theorem}: integer spin
particles are bosons and half-integer spin particles are
fermions. While many structural features of this formalism are present
already in the simpler non-relativistic models, the relativistic
models add some non-trivial features typical for quantum field
theories.

In the following we discuss a precise mathematical formulation of the
quantum field theory models described above. We emphasis the
functorial nature of this construction which makes manifest that it
also applies to other situations, e.g., where the bosons and fermions
are also coupled to a gravitational background, are considered in
other spacetime dimensions than $3+1$, etc.

\section{Second quantization: non-relativistic case}
Consider a quantum system of non-distinguishable particles where the
quantum mechanical description of one such particle is known. In
general, this one-particle description is given by a Hilbert space $h$
and one-particle observables and transformations which are
self-adjoint and unitary operators on $h$, respectively. The most
important observable is the Hamiltonian $H$. We will describe a
general construction of the corresponding many-body system.

\noindent {\bf Example.} As a motivating example we take the Hilbert space
$h=L^2(\R^3)$ of square-integrable functions $f(\vx)$, $\vx\in \R^3$, and the
Hamiltonian $H$ in \Ref{Sch}. A specific example for a unitary operator on $h$
is the gauge transformation $(Uf)(\vx) = \exp(\ii \chi(\vx)) f(\vx)$ with
$\chi$ a smooth, real-valued functions on $\R^3$.

In this example, the corresponding wave functions for $N$ identical such
particles are the $L^2$-functions $f_N(\vx_1,\ldots,\vx_N)$, $\vx_j\in\R^3$.
It is obvious how to extend one-particle observables and transformations to
such $N$-particle states: for example, the $N$-particle Hamiltonian
corresponding to $H$ in \Ref{Sch} is
\eq H_N = \sum_{j=1}^N \frac1{2m}(-\ii\nabla_{\vx_j} + e\vA(t,\vx_j) )^2
-e\phi(t,\vx_j) , \label{H} \eqend
and the $N$-particle gauge transformation $U_N$ is defined through
multiplication with $\prod_{j=1}^N \exp(\ii \chi(\vx_j))$.

For systems of indistinguishable particles it is enough to restrict to wave
functions which are even or odd under particle exchanges,
\eq f_N(\vx_1,\ldots,\vx_j,\ldots, \vx_k,\ldots, \vx_N) = \pm
f_N(\vx_1,\ldots,\vx_k,\ldots, \vx_j,\ldots, \vx_N) \eqend
for all $1\leq j<k\leq N$, with the upper and lower sign corresponding to
bosons and fermions, respectively (this empirical fact is usually taken as
postulate in non-relativistic many-body quantum physics). It is convenient to
define the zero-particle Hilbert space as $\C$ (complex numbers) and to
introduce a Hilbert space containing states with all possible particle
numbers: This so-called {\em Fock space} contains all states
\eq \left( \bma{c} f_0\\ f_1(\vx_1)\\ f_2(\vx_1,\vx_2) \\
f_3(\vx_1,\vx_2,\vx_3) \\ \vdots \ema \right) \label{Fock} \eqend
with $f_0\in \C$. The definition of $H_N$ and $U_N$ then naturally extends to
this Fock space; see below.

\bigskip
\noindent {\bf General construction.} The construction of Fock spaces and
many-particle observables and transformations just outlined in a specific
example is conceptually simple. An alternative, more efficient construction
method is to use {\em quantum fields} which we denote as $\psi(\vx)$ and
$\psi^\dag(\vx)$, $\vx\in\R^3$. They can be fully characterized by the
following (anti-) commutator relations,
\eq [\psi(\vx),\psi^\dag(\vy)]_\mp = \delta^3(\vx-\vy),\quad
[\psi(\vx),\psi(\vy)]_\mp = 0, \eqend
where $[a,b]_\mp\equiv ab\mp ba$, with the commutator and anti-commutators
(upper and lower signs) corresponding to the boson and fermion case,
respectively. It is convenient to `smear' these fields with one-particle wave
functions and define
\eq \psi(f) = \int_{\R^3} \dd^3 x\, \overline{f(\vx)} \psi(\vx) ,\quad
\psi^\dag (f) = \int_{\R^3} \dd^3 x\, \psi^\dag(\vx) f(\vx) \label{psi1}
\eqend
for all $f\in h$. Then the relations characterizing the field operators can be
written as
\eq [\psi(f),\psi^\dag(g)]_\mp = (f,g),\quad [\psi(f),\psi(g)]_\mp = 0 \quad
\forall f,g\in h \label{cacr} \eqend
where $(f,g)=\int_{\R^3}\dd^3 x\, \overline{f(\vx)} g(\vx)$ is the inner
product in $h$. The Fock space $\cF_\mp(h)$ can then be defined by postulating
that it contains a normalized vector $\Omega$ called {\em vacuum} such that 
\eq \psi(f)\Omega = 0 \quad \forall f\in h \eqend
and that all $\psi^{(\dag)}(f)$ are operators on $\cF_\mp(h)$ such that
$\psi^\dag(f)=\psi(f)^*$ where $*$ is the Hilbert space adjoint. Indeed, from
this we conclude that $\cF_\mp(h)$, as vector space, is generated by
\eq f_1\wedge f_2 \wedge \cdots \wedge f_N \equiv \psi^\dag(f_1) \psi^\dag(f_2)
\cdots \psi^\dag(f_N) \Omega \eqend
with $f_j\in h$ and $N=0,1,2,\ldots$, and that the Hilbert space inner product
of such vectors is
\eq \langle f_1\wedge f_2 \wedge \cdots \wedge f_N, g_1\wedge g_2 \wedge
\cdots \wedge g_M\rangle = \delta_{N,M} \sum_{P\in S_N} (\pm 1)^{|P|}
\prod_{j=1}^N (f_j,g_{Pj}) \eqend
with $S_N$ the permutation group, with $(+1)^{|P|}=1$ always and
$(-1)^{|P|}=+1$ and $-1$ for even and odd permutations, respectively.  The
many-body Hamiltonian $q(H)$ corresponding to the one-particle Hamiltonian $H$
now can be defined by the following relations,
\eq q(H)\Omega=0,\quad [q(H),\psi^\dag(f)] = \psi^\dag (Hf) \label{q} \eqend
for all $f\in h$ such that $Hf$ is defined. Indeed, this implies
\eq q(H) f_1\wedge f_2 \wedge \cdots \wedge f_N = \sum_{j=1}^N f_1\wedge f_2
\wedge \cdots \wedge (Hf_j) \wedge \cdots \wedge f_N \eqend
which defines a self-adjoint operator on $\cF_\mp(h)$, and it is easy to check
that this coincides with our down-to-earth definition of $H_N$ above.
Similarly the many-body transformation $Q(U)$ corresponding to a one-particle
transformation $U$ can be defined as
\eq
Q(U)\Omega=\Omega,\quad Q(U)\psi^\dag(f) = \psi^\dag(Uf) Q(U)\label{Q}
\eqend
for all $f\in h$, which implies 
\eq Q(U) f_1\wedge f_2 \wedge \cdots \wedge f_N = (Uf_1)\wedge (Uf_2) \wedge
\cdots \wedge (Uf_N)\eqend
and thus coincides with our previous definition of $U_N$.

While we presented the construction above for a particular example, it
is important to note that it actually does not make reference to what
the one-particle formalism actually is. For example, if we had a model
of particles on a space $\cM$ given by some `nice' manifold of any
dimension and with $M$ internal degrees of freedom, we would take
$h=L^2(\cM)\otimes \C^M$ and replace \Ref{psi1} by
\eq \psi(f) = \int_\cM \dd\mu(\vx) \sum_{j=1}^M \overline{f_j(\vx)}
\psi_j(\vx) \eqend
and its hermitian conjugate, with the measure $\mu$ on $\cM$ defining
the inner product in $h$, $(f,g)=\int\dd\mu(\vx) \sum_j
\overline{f_j(\vx)}g_j(\vx)$. With that, all formulas after \Ref{psi1}
hold true as they stand. {\em Given any one-particle Hilbert space $h$
with inner product $(\cdot,\cdot)$, observable $H$, and transformation
$U$, the formulas above define the corresponding Fock spaces
$\cF_\mp(h)$ and many-body observable $q(H)$ and transformation
$Q(U)$.} It is also interesting to note that this construction has
various beautiful general (functorial) properties: the set of
one-particle observables has a natural Lie algebra structure with the
Lie bracket given by the commutator (strictly speaking: $\ii$ times
the commutator, but we drop the common factor $\ii$ for
simplicity). The definitions above imply
\eq [q(A),q(B)] = q([A,B]) \label{qq} \eqend
for one-particle observables $A,B$, i.e., the above-mentioned Lie
algebra structure is preserved under this map $q$. In a similar
manner, the set of one-particle transformations has a natural group
structure preserved by the map $Q$,
\eq Q(U)Q(V) = Q(UV),\quad Q(U)^{-1} = Q(U^{-1}).  \label{QQ} \eqend
Moreover, if $A$ is self-adjoint, then $\exp(\ii A)$ is unitary, and one can
show that
\eq Q(\exp(\ii A)) = \exp(\ii q(A)).  \eqend
For later use we note that, if $\{ f_n\}_{n\in\Z}$ is some complete,
orthonormal basis in $h$, then operators $A$ on $h$ can be represented by
infinite matrices $(A_{mn})_{m,n\in\Z}$ with $A_{mn}= (f_m,Af_n)$, and
\eq q(A) = \sum_{m,n} A^{\phantom\dag}_{mn} \psi_m^\dag
\psi_n^{\phantom\dag}\label{qA1}\eqend
where $\psi^{(\dag)}_n = \psi^{(\dag)}(f_n)$ obey
\eq
[\psi^{\phantom\dag}_m,\psi_n^\dag]_\mp = \delta_{m,n}, \quad 
[\psi^{\phantom\dag}_m,\psi_n^\dag]_\mp = 0 \quad \label{psimn}
\eqend
for all $m,n$. We also note that, in our definition of $q(A)$, we made a
convenient choice of normalization, but there is no physical reason to not
choose a different normalization and define
\eq
q'(A) = q(A) -b(A) 
\eqend
where $b$ is some linear function mapping self-adjoint operators $A$ to real
numbers.  For example, one may wish to use another reference vector
$\tilde\Omega$ instead of $\Omega$ in the Fock space, and then would choose
$b(A)=\langle\tilde \Omega,q(A)\tilde \Omega\rangle$. Then the relation in
\Ref{qq} are changed to
\eq [q'(A),q'(B)] = q'([A,B]) + S_0(A,B) \label{tqq}
\eqend
where $S_0(A,B) = b([A,B])$. However, the $\C$-number term $S_0(A,B)$ in the
relations \Ref{tqq} is trivial since it can be removed by going back to
$q(A)$.

\bigskip

\noindent {\bf Physical interpretation.} The Fock space $\cF_\mp(h)$ is the
direct sum of subspaces of states with different particle numbers $N$,
\eq \cF_\mp (h) = \bigoplus_{N=0}^\infty h_\mp^{(N)} \eqend
where the zero-particle subspace $h_\mp^{(0)}=\C$ is generated by the vacuum
$\Omega$, and $h_\mp^{(N)}$ is the $N$-particle subspace generated by the
states $f_1\wedge f_2\wedge\ldots \wedge f_N$, $f_j\in h$. We note that
\eq \cN \equiv q(1) \eqend
is the {\em particle number operator}, $\cN F_N=NF_N$ for all $f_N\in
h_\mp^{(N)}$.  The field operators obviously change the particle
number: $\psi^\dag(f)$ increases the particle number by one (maps
$h_\mp^{(N)}$ to $h_\mp^{(N+1)}$), and $\psi(f)$ decreases it by
one. Since every $f\in h$ can be interpreted as one-particle state, it
is natural to interpret $\psi^\dag(f)$ and $\psi(f)$ as {\em creation}
and {\em annihilation operators}, respectively: they create and
annihilate one particle in the state $f\in h$. It is important to note
that, in the fermion case, \Ref{cacr} implies $\psi^\dag(f)^2=0$,
which is a mathematical formulation of the {\em Pauli exclusion
principle}: it is not possible to have two fermions in the same
one-particle state. In the boson case there is no such restriction.
Thus, even though the formalisms used to describe boson- and fermion
systems look very similar, they describe dramatically different
physics.

\bigskip

\noindent {\bf Applications.}  In our example, the many-body Hamiltonian
$\cH_0\equiv q(H)$ can also be written in the following suggestive form,
\eq \cH_0 = \int\dd^3 x\, \psi^\dag(\vx) (H\psi)(\vx) , \eqend
and similar formulas hold true for other observables and other Hilbert spaces
$h=L^2(\cM)\otimes \C^n$.  It is rather easy to solve the model defined by
such Hamiltonian: all necessary computations can be reduced to one-particle
computations. For example, in the static case where $\vA$ and $\phi$ are time
independent, a main quantity of interest in statistical physics is the free
energy
\eq \cE \equiv -\beta^{-1} \log \left( {\rm Tr} \left( \exp{(-\beta[\cH_0-\mu
\cN]) } \right) \right) \eqend
where $\beta>0$ here is the inverse temperature, $\mu$ the chemical potential,
and the trace over the Fock space $\cF_\mp(h)$. One can show that
\eq \cE = \pm {\rm tr}\left( \beta^{-1} \log(1\mp
\exp({-\beta[H-\mu]}))\right) \eqend
where the trace here is over the one-particle Hilbert space $h$. Thus, to
compute $\cE$, one only needs to find the eigenvalues of $H$.

It is important to mention that the framework discussed here is not only for
external field problems but can be equally well used to formulate and study
more complicated models with interparticle interactions.  For example, while
the model with the Hamiltonian $\cH_0$ above is often too simple to describe
systems in nature, it is easy to write down more realistic models, e.g., the
Hamiltonian
\eq \cH= \cH_0 + (e^2/2)\int\dd^3 x\int\dd^3 y\, \psi^\dag(\vx) \psi^\dag(\vy)
|\vx-\vy|^{-1} \psi(\vy)\psi(\vx) \eqend
describes electrons in an external electromagnetic field interacting
through Coulomb interactions. This illustrates an important point
which we would like to stress: the task in quantum theory is two-fold,
namely to {\em formulate} and to {\em solve} (exact of otherwise)
models. Obviously, in the non-relativistic case, it is equally simple
to formulate many-body models with and without inter-particle
interactions, and the latter only are simpler because they are easier
to solve: the two tasks of formulating and solving models can be
clearly separated. As we will see, in the relativistic case, even the
formulation of an external field problem is non-trivial, and one finds
that one cannot formulate the model without at least partially solving
it. This is a common feature of quantum field theories making them
challenging and interesting.

\section{Relativistic fermion and boson systems}
\label{sec_rel} 
We now generalize the formalism developed in the previous section to the
relativistic case.

\bigskip \noindent{\bf Field algebras and quasi-free representations.} In the
previous section we identified the field operators $\psi^{(\dag)}(f)$ with
particular Fock space operators. This is analog to identifying the operators
$p_j=-\ii \partial_{x_j}$ and $q_j=x_j$ on $L^2(\R^M)$ with the generators of
the Heisenberg algebra, as usually done. (We recall: the Heisenberg algebra is
the star algebra generated by $P_j$ and $Q_j$, $j=1,2,\ldots,M<\infty$, with
the well-known relations,
\eq [P_j,P_k] = -\ii \delta_{jk}, \quad [P_j,P_k]=[P_j,Q_k]=0, \quad
P_j^\dag = P^{\phantom\dag}_j,\quad Q_j^\dag = Q^{\phantom\dag}_j  \eqend
for all $j,k$.) Identifying the Heisenberg algebra with a particular
representation is legitimate since, as is well-known, all its irreducible
representations are (essentially) the same (this statement is made precise by
a celebrated theorem due to von Neumann).

However, in case of the algebra generated by the field operators
$\psi^{(\dag)}(f)$, there exist representations which are truly different
from the ones discussed in the last section, and to construct relativistic
external field problems such representations are needed. It is therefore
important to distinguish the fields as generators of an algebra from the
operators representing them. We thus define the {\em (boson or fermion) field
algebra} $\cA_\mp(h)$ over a Hilbert space $h$ as the star algebra generated
by $\Psi^\dag(f)$, $f\in h$, such that the map $f\to \Psi(f)$ is linear and
the relations
\eq {[}\Psi(f),\Psi^\dag(g){]}_\mp = (f,g),\quad {[}\Psi(f),\Psi(g){]}_\mp =
0,\quad \Psi^\dag(f)^\dag = \Psi(f) 
\label{cacr1} 
\eqend
are fulfilled for all $f,g\in g$, with $\dag$ the star operation in
$\cA_\mp(h)$.

The particular representation of this algebra discussed in the last section
will be denoted by $\pi_0$, $\pi_0(\Psi^{(\dag)}(f)=\psi^{(\dag)}(f)$.  Other
representations $\pi_{P_-}$ can be constructed from any projection operators
$P_-$ on $h$, i.e., any operator $P_-$ on $h$ satisfying
$P_-^*=P_-^2=P_-^{\phantom 2}$. Writing $\hat\psi^{(\dag)}(f)$ short for
$\pi_{P_-}(\Psi^{(\dag)}(f))$, this so-called {\em quasi-free representation}
is defined by
\eqa \hat\psi^\dag(f) = \psi^\dag(P_+f) + \psi(\overline{P_-f}) ,\quad 
\hat\psi(f) = \psi(P_+f) \mp \psi^\dag(\overline{P_-f}) \label{QF} \eqaend
where the bar means complex conjugation. It is important to note that, while
the star operation is identical with the Hilbert space adjoint $*$ in the
fermion case, we have
\eq \hat\psi(f)^\dag = \psi(Ff)^* \, \mbox{ with $F=P_+-P_-$ for bosons}
\label{F} 
\eqend
where $F$ is a grading operator, i.e., $F^*=F$ and $F^2=1$. We stress
that the `physical' star operation always is $*$, i.e., physical
observables $A$ obey $A=A^*$.

The present framework suggests to regard quantization as the procedure
which amounts to going from a one-particle Hilbert space $h$ to the
corresponding field algebra $\cA_\mp(h)$. Indeed, the Heisenberg
algebra is identical with the boson field algebra $\cA_-(\C^M)$ (since
the latter is obviously identical with the algebra of $M$ harmonic
oscillators), and thus conventional quantum mechanics can be regarded
as boson quantization in the special case where the one-particle
Hilbert space is finite dimensional. It is interesting to note that
`fermion quantum mechanics' $\cA_-(\C^M)$ is the natural framework for
formulating and studying lattice fermion and spin systems which play
an important role in condensed matter physics.

In the following we elaborate the naive interpretations of the relativistic
equations in \Ref{KG} and \Ref{D} as a quantum theory of one particle, and we
discuss why they are unphysical. For simplicity we assume that the
electromagnetic fields $\phi,\vA$ are time independent. We then show that
quasi-free representations as discussed above can provide physically
acceptable many-particle theories. We first consider the Dirac case which is
somewhat simpler.

\subsection{Fermions} 
\noindent {\bf One-particle formalism:} Recalling that $\ii\partial_t$ is the
energy operator, we define the Dirac Hamiltonian $\D$ by rewriting \Ref{D} in
the following form,
\eq \ii\partial_t \psi = \D\psi,\quad \D = (-\ii \nabla +
e\vA)\cdot \valpha + m \beta -e\phi . \eqend
This Dirac Hamiltonian is obviously is a self-adjoint operator on the
one-particle Hilbert space $\hD=L^2(\R^4)\otimes\C^4$, but, different
from the Schr\"odinger Hamiltonian in \Ref{Sch}, it is not bounded
from below: for any $E_0>-\infty$ one can find a state $f$ such that
the energy expectation value $(f,\D f)$ is less than $E_0$. This can
be easily seen for the simplest case where the external potential
vanishes, $\vA=\phi=0$. Then the eigenvalues of $\D$ can be computed
by Fourier transformation, and one finds
\eq E = \pm \sqrt{\vp^2 + m^2 },  \quad \vp\in\R^3.  \label{Epm} \eqend
Due to the negative energy eigenvalues we conclude that there is no ground
state, and the Dirac Hamiltonian thus describes an unstable system which is
physically meaningless.

To summarize: a (unphysical) one-particle description of relativistic fermions
is given by a Hilbert space $\hD$ together with a self-adjoint Hamiltonian
$\D$ unbounded from below. Other observables and transformations are given by
self-adjoint and unitary operators on $\hD$, respectively.

\noindent {\bf Many-body formalism:} We now explain how to construct a
physical many-body description from these data. To simplify notation
we first assume that $\D$ has a purely discrete spectrum (which can be
achieved by using a compact space). We then can label the
eigenfunctions $f_n$ by integers $n$ such that the corresponding
eigenvalues $E_n\geq 0$ for $n\geq 0$ and $E_n<0$ for $n<0$. Using the
naive representation of the fermion field algebra discussed in the
last section we get (we use the notation introduced in \Ref{qA1})
\eq
q(\D) = \sum_{n\geq 0} |E_n| \psi_n^\dag \psi_n^{\phantom\dag} 
-\sum_{n <0} |E_n| \psi_n^\dag \psi_n^{\phantom\dag}, 
\eqend
which is obviously not bounded from below and thus not physically
meaningful. However, $\psi_n^\dag \psi_n^{\phantom\dag} = 1 - \psi_n^{\phantom
\dag} \psi_n^{\dag}$, which suggests that we can remedy this problem by
interchanging the creation- and annihilation operators for $n<0$.  This is
possible: it is easy to see that
\eq \hat\psi^{\phantom\dag}_n \equiv \psi^{\phantom\dag}_n\quad \forall n\geq
0\\ \;\mbox{ and }\; \hat\psi^{\phantom\dag}_n \equiv \psi^\dag_n\quad \forall
n< 0 \eqend
provides a representation of the algebra in \Ref{psimn}.  We thus define
\eq \hat q(\D) \equiv \sum_{n\in\Z} E^{\phantom\dag}_n :\hat\psi_n^\dag
\hat\psi_n^{\phantom\dag} : \eqend
with the so-called {\em normal ordering} prescription
\eq :\psi_m^\dag \psi^{\phantom\dag}_n: \; \equiv \psi_m^\dag
\psi^{\phantom\dag}_n - \langle\Omega, \psi_m^\dag \psi^{\phantom\dag}_n\Omega
\rangle \label{nn} , \eqend
where we made use of the freedom of normalization explained after \Ref{psimn}
to eliminate unwanted additive constants. We get $q(\D) =
\sum_{n\in\Z}|E^{\phantom\dag}_n|\psi_n^{\dag}\psi_n^{\phantom\dag}$, which is
manifestly a non-negative self-adjoint operator with $\Omega$ as groundstate.
We thus found a physical many-body description for our model.  We now can
define for other one-particle observables,
\eq \hat q(A) \equiv \sum_{n\in\Z} A^{\phantom\dag}_{mn} :\hat\psi_m^\dag
\hat\psi_n^{\phantom\dag} : , \eqend
and by straightforward computations we obtain
\eq [\hat q(A),\hat q(B)] = \hat q([A,B]) + S(A,B) \label{qqS} \eqend
where $S(A,B) = \sum_{m<0}\sum_{n\geq 0}(A_{mn}B_{nm}- B_{mn}A_{nm})$, i.e.,
\eq S(A,B) = {\rm tr}\left(P_-AP_+BP_- - P_-BP_+AP_- \right) \label{S} \eqend
with $P_- =\sum_{n<0}f_n(f_n,\cdot)$ the projection onto the subspace spanned
by the negative energy eigenvectors of $\D$ and $P_+=1-P_-$.  One can show
that $\hat q(A)$ no longer is defined for {\em all} operators but only if
\eq P_-AP_+ \; \mbox{ and } \; P_+AP_- \; \mbox{ are Hilbert-Schmidt
operators} \label{HS} \eqend
(we recall that $a$ is a Hilbert-Schmidt operator if ${\rm tr}(a^*a)<\infty$).
The $\C$-number term $S(A,B)$ in \Ref{qqS} is often called {\em Schwinger
term}, and different from the similar term in \Ref{tqq} it now is non-trivial,
i.e., it no longer is possible to remove it be a redefinition $\hat q'(A)=\hat
q(A)-b(A)$. This Schwinger term is an example of an anomaly, and it has
various interesting implications.

In a similar manner, one can construct the many-body transformations $\hat
Q(U)$ of unitary operators $U$ on $h$ satisfying the very Hilbert-Schmidt
condition in \Ref{HS}, and one obtains
\eq \hat Q(U) \hat Q(V) = \chi(U,V) \hat Q(UV) \label{QQc} \eqend
with an interesting phase valued functions $\chi$. 

More generally, for any one-particle Hilbert space $h$ and Dirac Hamiltonian
$\D$, the physical representation is given by the quasi-free representation
$\pi_{P_-}$ in \Ref{QF} with $P_-$ the projection onto the negative energy
subspace of $\D$. The results about $\hat q$ and $\hat Q$ mentioned hold true
in any such representation. 

Thus the one-particle Hamiltonian $\D$ determines which representation one has
to use, and one therefore cannot construct the `physical' representation
without specific information about $\D$. However, not all these
representations are truly different: If there is a unitary operator $\cU$ on
the Fock space $\cF_+(h)$ such that
\eq \cU^* \pi_{P_-^{(1)}}(\psi^{(\dag)}(f)) \cU =
\pi_{P_-^{(2)}}(\psi^{(\dag)}(f)) \label{UE} \eqend
for all $f\in h$, then the quasi-free representations associated with
the different projections $P_-^{(1)}$ and $P_-^{(2)}$ are physically
equivalent: one could equally well formulate the second model using
the representation of the first. Two such quasi-free representations
are called {\em unitarily equivalent}, and a fundamental {\em theorem}
due to {\em Shale and Stinespring} states that two quasi-free
representations $\pi_{P_-^{(1,2)}}$ are unitarily equivalent if and
only if $P_-^{(1)}-P_-^{(2)}$ is a Hilbert-Schmidt operator (a similar
result holds true in the boson case).

\subsection{Bosons} 
\noindent {\bf One-particle formalism:} Similarly as for the Dirac
case, also the solutions of the Klein-Gordon equation in \Ref{KG} do
not define a physically acceptable one-particle quantum theory with a
ground state: the energy eigenvalues in \Ref{Epm} for $\vA=\phi=0$ are
a consequence the relativistic invariance and thus equally true for
the Klein-Gordon case. However, in this case there is a further
problem. To find the one-particle Hamiltonian one can rewrite the
second order equation in \Ref{KG} as a system of first order
equations,
\eq \ii \partial_t \Phi = \K\Phi,\quad \Phi = \left(\bma{c}
\psi\\\pi^\dag\ema\right), \quad \K = \left(\bma{rr} C & \ii \\ -\ii B^2 & C
\ema \right) \label{K} \eqend
with
\eq B^2 \equiv ( -\ii \nabla + e\vA)^2 + m^2 
,\quad C\equiv -e\phi .  \eqend
Thus one sees that the natural one-particle Hilbert space for the
Klein-Gordon equation is $\hKG=L^2(\R^3)\otimes \C^2$; here and in the
following we identify $\hKG$ with $h_0\oplus h_0$, $h_0=L^2(\R^3)$,
and use a convenient $2\times 2$ matrix notation naturally associated
with that splitting.  However, the one-particle Hamiltonian is not
self-adjoint but rather obeys
\eq \K^* = J\K J,\quad J\equiv \left(\bma{rr} 0&-\ii\\ \ii & 0
\ema\right) \label{J} \eqend
with $*$ the Hilbert space adjoint. It is important to note that $J$ is a
grading operator. Thus, we can define a sequilinear form
\eq (f,g)_J \equiv (f,Jg) \quad \forall f,g\in \hKG , \eqend
with $(\cdot,\cdot)$ the standard inner product, and \Ref{J} is
equivalent to $\K$ being self-adjoint with respect to this
sesquilinear form; in this case we say that $\K$ is {\em
$J$-self-adjoint}.  Thus, in the Klein-Gordon case, this sesquilinear
form takes the role of the Hilbert space inner product and, in
particular, not $(\Phi,\Phi)$ but $(\Phi,\Phi)_J$ is preserved under
time evolution.  However, different from $\Phi^\dag \Phi$, $\Phi^\dag
J\Phi$ is not positive definite, and it is therefore not possible to
interpret it as probability density as in conventional quantum
mechanics. For consistency one has to require that one-particle
transformations $U$ are unitary with respect to $(\Phi,\Phi)_J$, i.e.,
$U^{-1}=JUJ$. We call such operators {\em $J$-unitary}.

To summarize: a (unphysical) one-particle description of relativistic
bosons is given by a Hilbert space of the form $\hKG=h_0\oplus h_0$,
the grading operator $J$ in \Ref{J}, and a $J$-self-adjoint
Hamiltonian $\K$ of the form as in Eq.\ \Ref{K} where $B\geq 0$ and
$C$ are self-adjoint operators on $h_0$. Other observables and
transformations are given by $J$-self-adjoint and $J$-unitary
operators on $\hKG$, respectively.

\noindent {\bf Many-body formalism:} We first consider the quasi-free
representation $\pi_{P_-^{(0)}}$ of the boson field algebra $\cA_-(\hKG)$ so
that the grading operator in \Ref{F} is equal to $J$, i.e., $P_-^{(0)} =
(1-J)/2$. Writing $\pi_{P_-^{(0)}}(\Psi^{(\dag)}(f))= \psi^{(\dag)}(f)$ one
finds that
\eq q(A)^* = q(JAJ),\quad Q(U)^* = Q(JU^*J) , \eqend
and thus $J$-selfadjoint operators and $J$-unitary operators are mapped to
proper observables and transformations. In particular, $q(\K)$ is a
self-adjoint operator, which resolves one problem of the one-particle
theory. However, $q(\K)$ is not bounded from below, and thus $\pi_{P_-^{(0)}}$
is not yet the physical representation.

The physical representation can be constructed using the operators
\eq T = \frac1{\sqrt2} \left( \bma{rr} B^{1/2} & \ii B^{-1/2} \\ B^{1/2} & \ii
B^{-1/2} \ema\right) ,\quad F = \left( \bma{rr} 1 & 0 \\ 0 &  -1\ema\right)
\eqend
(for simplicity we restrict ourselves to the case $C=0$ and $B>0$; we use of
the calculus of self-adjoint operators here) with the following remarkable
properties,
\eq T^{-1} = JT^* F,\quad T\K T^{-1} = \left( \bma{rr} B & 0 \\ 0 &
-B\ema\right)\equiv \KK.  \eqend
One can check that 
\eq \hat\psi^\dag(f) \equiv \psi^\dag(Tf),\quad \hat\psi(f) \equiv \psi(T^{-1}
f),\quad \eqend
is a quasi-free representation $\pi_{P_-}$ of $\cA_-(\hKG)$ with
$P_-=(1 -F)/2$.  With that the construction of $\hat q$ and $\hat Q$
is very similar to the fermion case described above (the crucial
simplification is that $\KK$ and $F$ now are diagonal). In particular,
$\hat q(K)$ is a non-negative operator with the ground state $\Omega$,
and $\hat q(A)$ and $\hat Q(U)$ is self-adjoint and unitary for every
one-particle observable $A$ and transformation $U$, respectively. One
also gets relations as in \Ref{qqS} and \Ref{QQc}.

\section{Further reading} 
The impossibility to construct relativistic quantum mechanical models
played an important role in the early history of quantum field theory,
as beautifully discussed in \cite{Weinberg}, Chapter~1.

The abstract formalism of quasi-free representations of fermion and
boson field algebras was developed in many papers; see e.g.\
\cite{R,GL,EL} for explicit results on $\hat Q$ and $\chi$. A nice
textbook presentation with many references can be found in \cite{NCG},
Chapter~13 (this chapter is rather self-contained but mainly
restricted to the fermion case).

Based on the Shale-Stinespring theorem there has been considerable
amount of work to investigate whether the quasi-free representations
associated with different external electromagnetic fields
$\psi_1,\vA_1$ and $\psi_2,\vA_2$ are unitarily equivalent, if and
which time dependent many-body Hamiltonians exist etc.; see
\cite{NCG}, Chapter 13 and references therein.

The infinite dimensional Lie $g_2$ of Hilbert space operators
satisfying the condition in \Ref{HS} is an interesting infinite
dimensional Lie algebra with a beautiful representation theory. This
subject is closely related to conformal field theory; see e.g.\
\cite{Kac} for a textbook presentation and \cite{CR} for a detailed
mathematical account within the framework described by us.

It turns out that the mathematical framework discussed in
Section~\ref{sec_rel} is sufficient for constructing fully interacting
quantum field theories, in particular Yang-Mills gauge theories, in
1+1 but not in higher dimensions. The reason is that, in 3+1
dimensions, the one-particle observables $A$ of interest do not obey
the Hilbert-Schmidt condition in \Ref{HS} but only the weaker
condition,
\eq {\rm tr}(a^*a)^n<\infty ,\quad a=P_\mp A P_\pm .  \eqend
with $n=2$, and the natural analog of $g_2$ in 3+1 dimensions thus
seems to be the Lie algebra $g_{2n}$ of operators satisfying this
condition with $n=2$.  Various results on the representation theory of
such Lie algebras $g_{2n>2}$ have been developed; see
\cite{Mickelsson} where also various interesting relations to infinite
dimensional geometry are discussed.

As mentioned, the Schwinger term $S(A,B)$ in \Ref{S} is an example of
an anomaly. Mathematically it is a non-trivial 2-cocycles of the Lie
algebra $g_2$, and analogs for the groups $g_{2n>2}$ have been
found. These cocycles provide a natural generalization of anomalies
(in the meaning of particle physics) to operator algebras. They not
only shed some interesting light on the latter, but also provide a
link to notions and results from non-commutative geometry; see e.g.\
\cite{NCG}. We believe that this link can provide a fruitful driving
force and inspiration to find ways to deepen our understanding of
quantum Yang-Mills theories in 3+1 dimensions \cite{L}.

\bigskip

\section*{Keywords}

Conformal field theory\\ 
anomalies\\
Noncommutative geometry\\ 
Dirac operators\\
Determinants of differential operators\\

\end{document}